\documentclass[numbered]{trbunofficial}
\usepackage{graphicx}
\usepackage{float}
\usepackage{booktabs}
\usepackage{subfigure}
\usepackage{amsmath,amssymb} 
\usepackage{bm}
\usepackage{changepage}
\usepackage[hidelinks]{hyperref}

\title{Reinforcement Learning from Human Feedback for Lane Changing of Autonomous Vehicles in Mixed Traffic}
\author{Yuting Wang, Lu Liu, Maonan Wang and Xi Xiong}

\begin{document}
\maketitle
\newcommand\blfootnote[1]{%
\begingroup
\renewcommand\thefootnote{}\footnote{#1}%
\addtocounter{footnote}{-1}%
\endgroup
}
\blfootnote{This work was supported in part by NSFC Project 72371172 and Fundamental Research Funds for the Central Universities.

Y. Wang, L. Liu and X. Xiong are with the Key Laboratory of Road and Traffic Engineering, Ministry of Education, Tongji University, Shanghai, China.  M. Wang is with the School of Science and Engineering, the Chinese University of Hong Kong, Shenzhen, China (Emails: wangyuting321@tongji.edu.cn, luliu0720@tongji.edu.cn, maonanwang@link.cuhk.edu.cn, xi\_xiong@tongji.edu.cn,)}


\titleformat{name=\section,numberless}[block]{\filcenter\Large\bfseries}{}{0em}{}
\vspace{-5em}
\begin{adjustwidth}{3em}{3em}
\section*{Abstract}

The burgeoning field of autonomous driving necessitates the seamless integration of autonomous vehicles (AVs) with human-driven vehicles, calling for more predictable AV behavior and enhanced interaction with human drivers. Human-like driving, particularly during lane-changing maneuvers on highways, is a critical area of research due to its significant impact on safety and traffic flow. Traditional rule-based decision-making approaches often fail to encapsulate the nuanced boundaries of human behavior in diverse driving scenarios, while crafting reward functions for learning-based methods introduces its own set of complexities. This study investigates the application of Reinforcement Learning from Human Feedback (RLHF) to emulate human-like lane-changing decisions in AVs. An initial RL policy is pre-trained to ensure safe lane changes. Subsequently, this policy is employed to gather data, which is then annotated by humans to train a reward model that discerns lane changes aligning with human preferences. This human-informed reward model supersedes the original, guiding the refinement of the policy to reflect human-like preferences. The effectiveness of RLHF in producing human-like lane changes is demonstrated through the development and evaluation of conservative and aggressive lane-changing models within obstacle-rich environments and mixed autonomy traffic scenarios. The experimental outcomes underscore the potential of RLHF to diversify lane-changing behaviors in AVs, suggesting its viability for enhancing the integration of AVs into the fabric of human-driven traffic.

\hfill\break%
\noindent\emph{Keywords}: 
Autonomous Vehicles, Human-like Driving, Reinforcement Learning from Human Feedback, Proximal Policy Optimization

\end{adjustwidth}
\section{Introduction}

In recent years, both academia and industry have shown great interest in the development of autonomous driving, a technology \cite{world2023helmets} that can radically liberate drivers while providing new ideas for solving the problems of road safety, ineffective use of road infrastructure and pollutant emissions. However, as it will take decades for autonomous vehicles (AVs) to become fully ubiquitous, it is inevitable that they will share the road with human-driven vehicles for a long time \cite{hang2020human, xiong2024approximate}. In the open road testing of AVs in California, accidents related to AVs have been dominated by rear-end collisions by human drivers, accounting for 57\% of the accidents. The heterogeneity between the behavior of AVs and human drivers is the main cause of the accidents \cite{zhang2023human}. In order to improve the behavioral predictability of AVs, adopting the concept of human-like driving has become an important research direction \cite{xu2020learning}.
Besides, lane-changing behavior is one of the most frequent and inherently risky maneuvers in fundamental highway driving behaviors. It significantly influences both the safety and the passing efficiency of vehicular traffic. According to the U.S. Highway Traffic Safety Administration (NHTSA) statistics \cite{national2008national}, up to 27\% of road accidents are caused by unreasonable lane changing behavior. Compared to human-driven vehicles, AVs are able to obtain timely information about the movement of surrounding vehicles, and improve road safety by precisely controlling position, speed, and shorter reaction time. Therefore, it is of great significance to study in depth the lane changing behavior of AVs.

A large number of studies have applied predefined rule-based models to solve lane-changing decision-making problems in the presence of vehicle interactions. Yang et al. \cite{yang1996microscopic}, Hidas et al. \cite{hidas2005modelling}, and Kesting et al. \cite{kesting2007general} modelled the lane-changing process based on certain assumptions. This approach oversimplifies the real lane-changing problem. Although vehicles may be better at lane-changing in predefined situations, it is difficult to adapt to the dynamic changes in traffic flow in real complex environments.
To address the limitations of traditional rule-based approaches, recent research has applied reinforcement learning (RL) for lane-changing decisions of AVs \cite{han2023leveraging}. Chen et al. \cite{chen2020autonomous}, and Krasowski et al. \cite{krasowski2020safe} combined RL to improve the safety and stability of the vehicle lane change process. However, the reward function, an important component of RL, has traditionally been used in a way that requires human design and is difficult to cover all aspects in complex scenarios.
In terms of realising human-like autonomous driving, existing methods that rely on driving style differentiation and related feature extraction are limited by rule setting. It is hard to represent all aspects of human preferences and can not comprehensively cover individual human needs. These methods often use real-world driving data and driving simulator data, which are also difficult and expensive to obtain.

To address these challenges, we employ Reinforcement Learning with Human Feedback (RLHF) to capture human driving habits, using human feedback as a learning signal to help AVs understand complex human preferences. In this paper, we focus on lane-changing decisions for AVs. Specifically, in order to speed up the training process, we first use Proximal Policy Optimization (PPO) to pre-train a vehicle lane-changing decision model, which also serves as the base model in our experiment. Instead of giving specific ratings, we then asked participants to compare different video clips of trajectories obtained by pre-trained models controlling vehicle lane changing. Participants evaluate the behaviour of an agent based on how well their lane-changing style matches their preferences. We use their preferences to fit a reward function which can predict rewards based on the interaction between the agent and the environment. Finally, we use the PPO algorithm for fine tuning to get different styles of RLHF models. This method avoids the difficulty of not being able to quantify the reward model in some special cases, improves the efficiency of RL and increases the interpretability of the decision process. Moreover, we can increase the economy of collecting feedback and meet the individual needs of passengers.

The main contributions of this paper include:
(i) formulating the human-like lane-changing problem as a Markov decision process (MDP);
(ii) utilizing the Reinforcement Learning from Human Feedback (RLHF) algorithm to fine-tune lane-changing decision-making for policy improvement;
(iii) validating the effectiveness of RLHF in obstacle avoidance and mixed autonomy scenarios within a simulation platform.

The rest of the paper is organized as follows.
In Section~\ref{section:realted_work}, we provide a literature review of lane-changing decisions of AVs and RLHF methods.
In Section~\ref{section:methods}, we propose the RLHF framework for lane changing decisions
In Section~\ref{section:experiments}, we validate the effectiveness of the RLHF method based on SUMO platform.
In Section~\ref{section:conclusion}, we summarize the conclusions and discuss future work.
\section{Related Works} \label{section:realted_work}

In the realm of autonomous vehicle lane-changing decision-making, methodologies are generally classified into rule-based and learning-based strategies. Rule-based systems utilize predefined conditions set by humans to determine the appropriateness of a lane change. Traditional models for lane-changing intentions typically prioritize factors such as higher travel speeds or more significant gaps between vehicles. Gipps \textit{et al.} introduced a set of criteria including a minimum safe distance model and a mechanism for avoiding obstacles during lane changes on multi-lane urban roads, which incorporated various driving environment factors \cite{gipps1986model}. Similarly, Naranjo \textit{et al.} employed a fuzzy logic controller designed to replicate human decision-making in lane changes, taking into account relative distances and velocities of the subject vehicle and surrounding traffic \cite{naranjo2008lane}. While rule-based systems have demonstrated effectiveness in situations that adhere to their predefined rules, they fall short in adapting to the unpredictable nature of real-world driving conditions and varying driver behaviors. This limitation has led to the exploration of learning-based approaches that aim to enhance decision-making in real-time and multifaceted traffic scenarios.

Transitioning from rule-based to learning-based methods, machine learning offers the ability to adapt to complex traffic environments through analysis of real-world data. These methods excel in discerning intricate patterns and adjusting to diverse conditions, which may enhance the precision of decision-making processes as they assimilate new information. For example, Liu \textit{et al.} presented a Support Vector Machine (SVM) model that integrates factors like benefits, safety, and driver tolerance to tailor lane change decisions to individual driving styles \cite{liu2019novel}. Similarly, Bi \textit{et al.} developed a data-driven Random Forest model that enables the consideration of multiple gap opportunities in the target lane for merging \cite{bi2016data}. RL-based method, a subset of machine learning, is particularly beneficial for lane-changing research due to its ability to derive optimal strategies through direct environmental interaction \cite{wang2024traffic}. Zhou \textit{et al.} constructed an innovative multi-agent advantage actor-critic network model, enhancing efficiency and scalability through parameter sharing and a well-crafted reward function \cite{zhou2022multi}. In a different approach, Wang \textit{et al.} implemented a quadratic function as the Q-function approximator in their RL model, with coefficients learned via neural networks, and a reward function emphasizing yaw rate, acceleration, and lane changing duration to encourage smooth and efficient maneuvers \cite{wang2018reinforcement}. Additionally, Li \textit{et al.} utilized deep RL to devise a lane-changing strategy aimed at minimizing expected risk \cite{li2022decision}, while Wang \textit{et al.} considered both individual vehicle delay and overall traffic flow efficiency in their reward function design \cite{wang2019cooperative}. Although these learning-based approaches are proficient in directing vehicles with commendable efficiency and safety, they have not yet mastered the emulation of human-like lane-changing behaviors.

Building upon rule-based and learning-based methods, the integration of human factors into lane-changing algorithms for AVs has been an emerging focus aimed at achieving human-like decision-making. Some researchers have incorporated elements such as real driving data and diverse driving styles to enhance the human-likeness of AV behavior. Zhu \textit{et al.} employed a reward function within their RL framework that penalizes deviations from real-world driving data, aligning the lane-changing policy closer to human behavior \cite{zhu2018human}. Similarly, Gu \textit{et al.} introduced a lane-changing decision (LCD) model grounded in actual on-road driving data, endowing AVs with the capability to mirror human decision-making processes \cite{gu2020novel}. Further efforts by Kuefler \textit{et al.} utilized Generative Adversarial Imitation Learning (GAIL) to model human driving patterns on highways, with expert input derived from real driver trajectories within the NGSIM dataset \cite{kuefler2017imitating}. Huang \textit{et al.} applied maximum entropy inverse reinforcement learning (IRL) to extract personalized reward functions for individual drivers, also leveraging the NGSIM dataset for a tailored approach \cite{huang2021driving}. Additionally, Hang \textit{et al.} addressed noncooperative decision-making in driving by applying Nash equilibrium and Stackelberg game theory, considering varied driving styles and the intricacies of social interactions \cite{hang2020human}.

Recent advancements have seen the application of RLHF \cite{ouyang2022training}, a technique that trains reward functions based on human preferences through supervised learning methods. This approach circumvents the challenge of manually crafting reward functions in complex scenarios and enhances the AI system's alignment with human preferences. Initially applied to fine-tune large language models, RLHF has improved the human-like quality of responses in terms of linguistic style and ethical considerations \cite{peng2023instruction}. However, its application in transportation research remains limited. Cao \textit{et al.} proposed an RLHF-based framework, TrafficRLHF, to refine the authenticity of traffic models by capturing the subtleties of human preferences and harmonizing various traffic simulation models \cite{cao2023reinforcement}. RLHF holds significant promise for supporting RL agents in making decisions that are more closely aligned with human behavior, providing clearer guidance with minimal human input. This paper aims to extend the application of RLHF to the traffic domain, specifically for lane-changing scenarios, to facilitate AVs in achieving decision-making processes that resonate with human preferences and driving patterns.
\section{Methodology} \label{section:methods}

\subsection{Framework Overview}

\begin{figure}[!ht]
  \centering
  \includegraphics[width=0.99\textwidth]{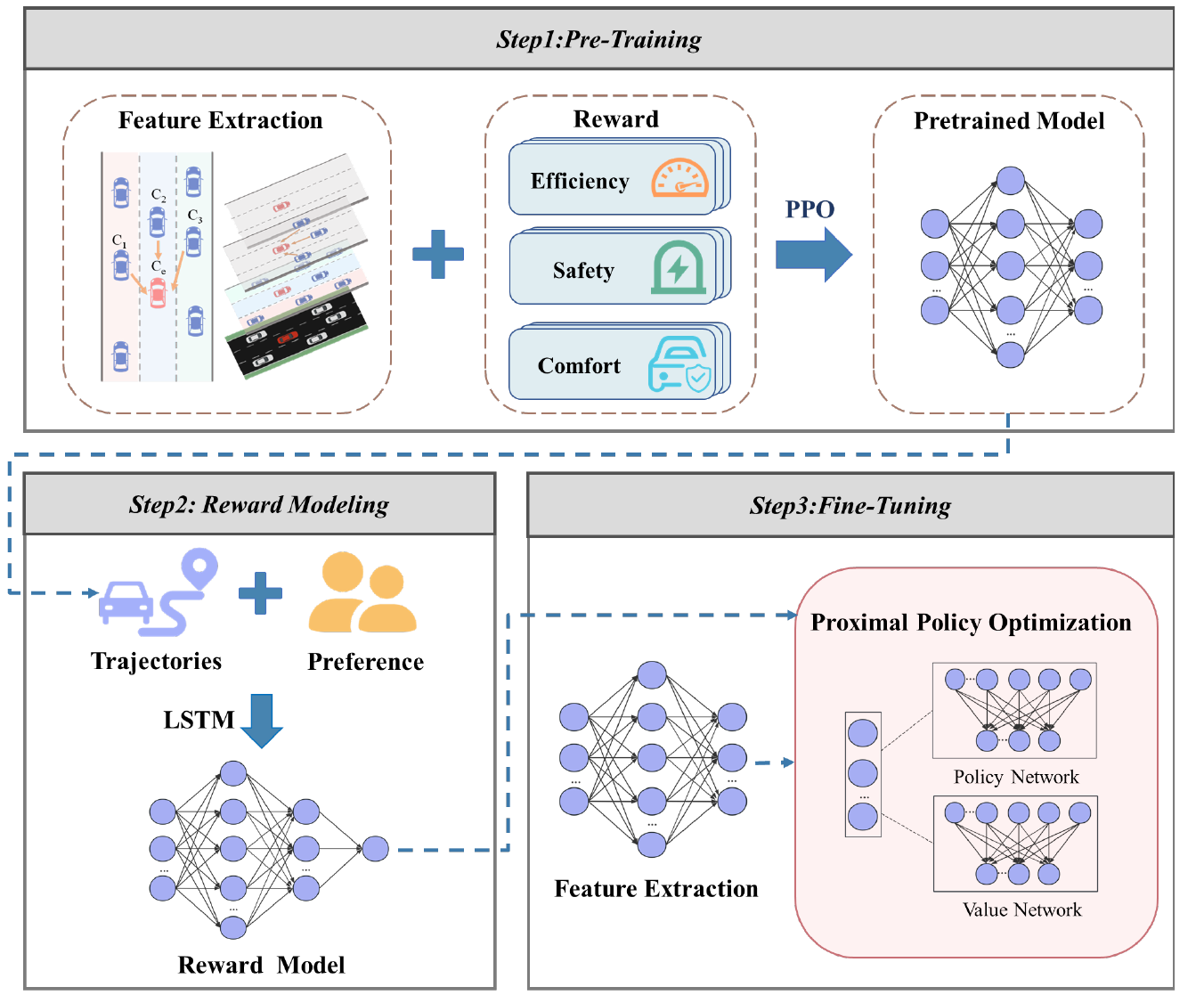}
  \caption{The overall framework of the lane changing model using RLHF.} 
  \label{fig:RLHF}
\end{figure}

As depicted in Figure~\ref{fig:RLHF}, a three-part framework is introduced for the enhancement of lane-changing maneuvers, incorporating human-like decision-making factors. Initially, the Proximal Policy Optimization (PPO) algorithm is employed to train the RL agent, enabling it to make lane change decisions. The agent is capable of processing data from surrounding vehicles and lane-specific information to execute lane changes. To ensure the initial policy demonstrates satisfactory performance, a reward function is meticulously crafted, which emphasizes driving efficiency, safety, and comfort. 

Following the pre-training phase, trajectory segments generated by the RL models are evaluated by human assessors. These segments are labeled based on the assessors' preferences. A reward model is subsequently developed to align with the human feedback. This model is instrumental in adjusting the initial policy. The refined policy, also achieved through the PPO algorithm, aims to replicate human-like preferences in the context of lane-changing maneuvers. The final model is a testament to the efficacy of integrating human feedback into the fine-tuning process, thereby facilitating the adoption of humanistic driving styles by AVs.

\subsection{Pre-training the Lane Change Model as an MDP}

In this subsection, the methodology for the pre-training phase of the lane change model is delineated. The pre-training enables the RL agent to acquire fundamental driving competencies, which include maintaining lane integrity, executing lane changes, and modulating velocity within a controlled simulation environment. The proficiency developed during this phase is crucial for the autonomous vehicle's ability to operate safely and avoid collisions with nearby traffic, as well as to perform lane changes with efficiency. In the forthcoming phase of strategy refinement, the parameters governing feature extraction will be held constant, while adjustments will be made solely to the strategy component to incorporate diverse driving behaviors. This approach ensures a stable foundation upon which nuanced driving styles can be superimposed.

We frame the lane-changing behavior of AVs within the MDP framework. Here, the AV uses sensor data to assess its immediate surroundings, including the position and movement of other vehicles, to gain a full understanding of the traffic context. The MDP is characterized by the tuple $(S, A, T, R, \gamma)$, where $S$ represents the set of states corresponding to global traffic conditions, which the AV is presumed to fully observe. $A$ defines the set of actions the AV can take, contingent on the current state. The state transition probability function, $T: S \times A \times S \rightarrow [0, 1]$, determines the likelihood of transitioning from one state to another following a specific action by the AV. The reward function, $R: S \times A \rightarrow \mathbb{R}$, assigns immediate value to actions taken in particular states. Finally, $\gamma \in [0,1]$ is the discount factor that weighs immediate against future rewards, ensuring a balanced approach to decision-making.

At each discrete time step $t$, the AV observes $\bm{o}_i^t \in O_i$, leading to a state $\bm{s}^t$ and making an action $\bm{a}^t$, which results in a transition to a new state $\bm{s}^{t+1}$ as per the transition function $T$. Concurrently, the AV receives a reward $r^t$, computed by the reward function $R$, for the action taken. The goal within the MDP framework is to discover an optimal policy $\pi^\ast: S \rightarrow A$ that maximizes the total expected reward over time. The value function $V^\ast(\bm{s})$ for a state $\bm{s}$ under the optimal policy is defined by the Bellman equation, this formulation lays the groundwork for subsequent specific settings and the integration of human feedback into the lane-changing policy:

\begin{equation}
    V^\ast(\bm{s}) 
    = 
    \max_{\bm{a} \in A} \left[ R(\bm{s},\bm{a}) + \gamma \sum_{\bm{s}' \in S} T(\bm{s},\bm{a},\bm{s}') V^\ast(\bm{s}') \right].
\end{equation}

The state representation within the MDP for managing lane change maneuvers is specified as follows. At each discrete time step $t$, the AV agent collects kinematic data pertaining to itself and the surrounding traffic. This data includes the current state of the ego vehicle, the nearest leading vehicle in the target lane, left lane and right lane of the ego vehicle, and aggregated lane-level information. The state observation for the AV is defined by the equation:

\begin{equation} \label{Eq_state}
S = (
    \underbrace{v_e, a_e}_{\text{ego vehicles}}, \;
    \underbrace{v_1, a_1, \Delta y_{1}, v_2, a_2, \Delta y_{2}, v_3, a_3, \Delta y_{3}}_{\text{leading vehicles}}, \quad
    \underbrace{n_{1}, \bar{v}_{1}, \ldots, n_{K}, \bar{v}_{K}}_{\text{lane}}
    ),
\end{equation}
where $K$ denotes the total number of lanes. The variables $v_i$ and $a_i$ represent the velocity and acceleration of the leading vehicle, while $\Delta y_{i}$ denotes the lateral displacement of the same vehicle relative to the AV. In the context of lane-level data, $n_j$ indicates the number of vehicles in the $j$-th lane, and $\bar{v}_{j}$ is the average speed of traffic within that lane. In instances where the number of leading vehicles is fewer than three, the absent values are substituted with zeros. 

\begin{figure}[!ht]
  \centering
  \includegraphics[width=0.8\textwidth]{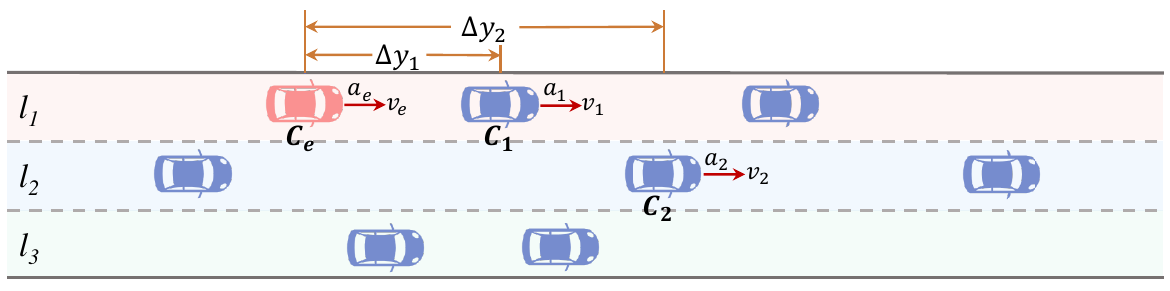}
  \caption{Illustration of a Three-Lane Vehicle Lane-Change Scenario.} 
  \label{fig:state_space}
\end{figure}

Figure~\ref{fig:state_space} illustrates the state space in a three-lane scenario, where vehicles $C_1$ and $C_2$ are positioned ahead of the AV in lanes $l_1$ and $l_2$ respectively. Since the left lane of $l_1$ does not exist, the corresponding values for $v_3$ and $a_3$ are set to zero, while $\Delta y_{3}$ is set to a fairly large number. This representation captures the necessary situational context to inform the AV's decision-making process for lane changes.

In the pre-training phase of lane-changing model, the agent selects an action $a$ from a predefined set: $\{a_{\text{left}}, a_{\text{right}}, a_{\text{keep}}, a_{\text{acc}}, a_{\text{dec}}\}$. These actions represent lane-changing to the left, lane-changing to the right, maintaining the current lane, accelerating, and decelerating, respectively.

The reward function $r$ in our initial module is designed to balance three key aspects of driving: efficiency, comfort, and safety. The total reward $r$ for a single action is the sum of the individual rewards for these aspects:

\begin{equation} \label{Eq.total_reward}
    r = r_s + r_{lc} + r_{e}.
\end{equation}

Safety is prioritized above all, with a substantial penalty assigned for collisions to prevent the AV from coming into contact with surrounding vehicles:

\begin{equation} \label{Eq.safety_reward}
     r_s =  
     \begin{cases}
      -C_s, & \text{if a collision occurs,} \\
      0, & \text{otherwise.}
    \end{cases}
\end{equation}

The comfort metric focuses on the frequency and necessity of lane changes, penalizing unnecessary or excessive maneuvers to discourage erratic behavior:

\begin{equation} \label{Eq.comfort_reward}
     r_{lc} =  
     \begin{cases}
      -C_{lc}, & \text{if a lane change is made,} \\
      0, & \text{otherwise.}
    \end{cases}
\end{equation}

The efficiency reward $r_e$ is given for maintaining a high speed without exceeding the legal speed limit. Let $v$ be the vehicle's current speed, $v_{max}$ the maximum permitted speed, and $v_{min}$ the minimum acceptable speed for the controlled vehicle:

\begin{equation} \label{Eq.efficiency_reward}
     r_{e} =  
     \begin{cases}
      (v - v_{min})/(v_{max} - v_{min}), & \text{if } v < v_{min}, \\
      (v_{max} - v)/(v_{max} - v_{min}), & \text{if } v > v_{max}, \\
      v/v_{max}, & \text{otherwise.}
    \end{cases}
\end{equation}

With safety as the foundation, the AV is tasked with striking an optimal balance between comfort and efficiency as dictated by the reward function. This balance ensures that lane changes are executed in a manner that is both appropriate and efficient.

\begin{figure}[!ht]
  \centering
  \includegraphics[width=1.0\textwidth]{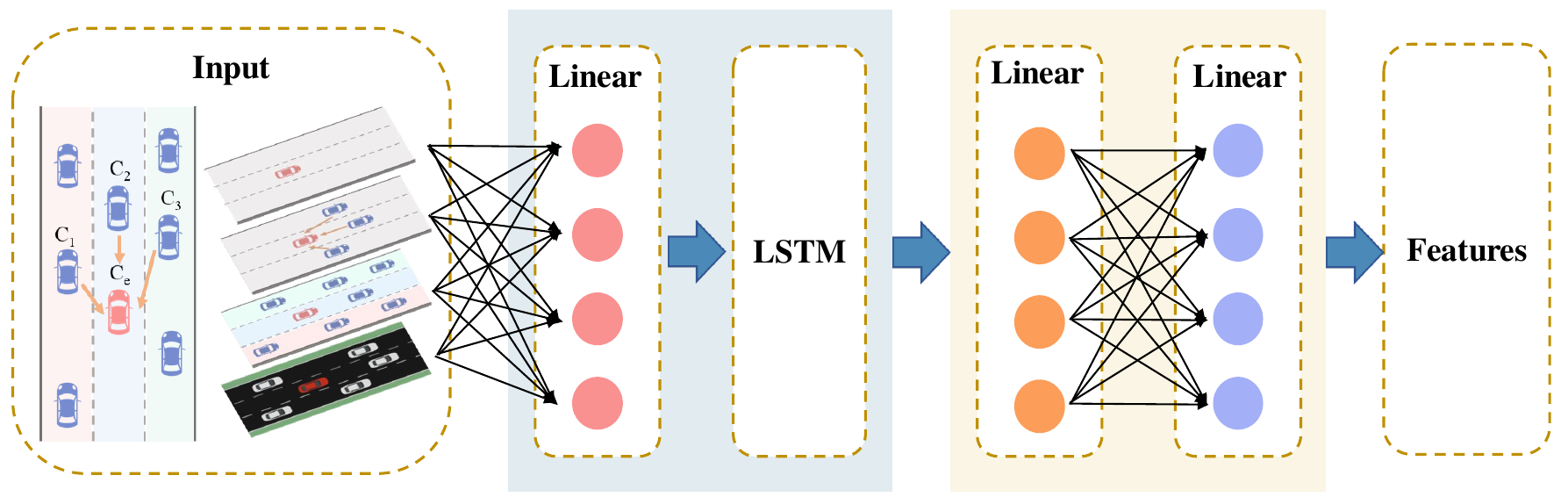}
  \caption{The architecture diagram of the feature extraction network. The blue color indicates the range of parameters that need to be fixed during the subsequent fine-tuning process.} 
  \label{fig:Feature}
\end{figure}

During the pre-training process, we construct a feature extraction network incorporating Long Short-Term Memory (LSTM) as illustrated in Figure~\ref{fig:Feature}, to extract features from the input observations. We transform the original observations into feature vectors with enhanced representational capabilities, thereby improving the agent's performance and training efficiency.

\subsection{Human Feedback Integration and Policy Adjustment}

After pre-training, we established a baseline policy capable of executing lane changes. This section aims to refine the policy to better emulate human-like lane-changing behavior. We divide this process into three stages: \textit{Collection of Human-Driven Behavioral Data}, \textit{Training of the Reward Model from Human Assessments}, and \textit{Policy Refinement for Human-Like Decision Making}. Initially, we gather human feedback by presenting scenarios generated by the pre-trained network to human evaluators. Subsequently, we employ supervised learning methods to construct a reward model that reflects these human judgments. Finally, we optimize the policy, adjusting the autonomous vehicle's decision-making process to align with the preferences indicated by the human-informed reward model.

\subsubsection{Collection of Human-Driven Behavioral Data}

In this stage, we collect human feedback data to understand and quantify human preferences regarding lane-changing behavior. To facilitate this, we employ the pre-trained model to generate a series of trajectories $\{\tau^1,\tau^2,...,\tau^k\}$ by simulating the agent's movement on a virtual road environment. The agent operates within this environment based on the baseline policy obtained from pre-training.

\begin{figure}[!ht]
  \centering
  \includegraphics[width=1.0\textwidth]{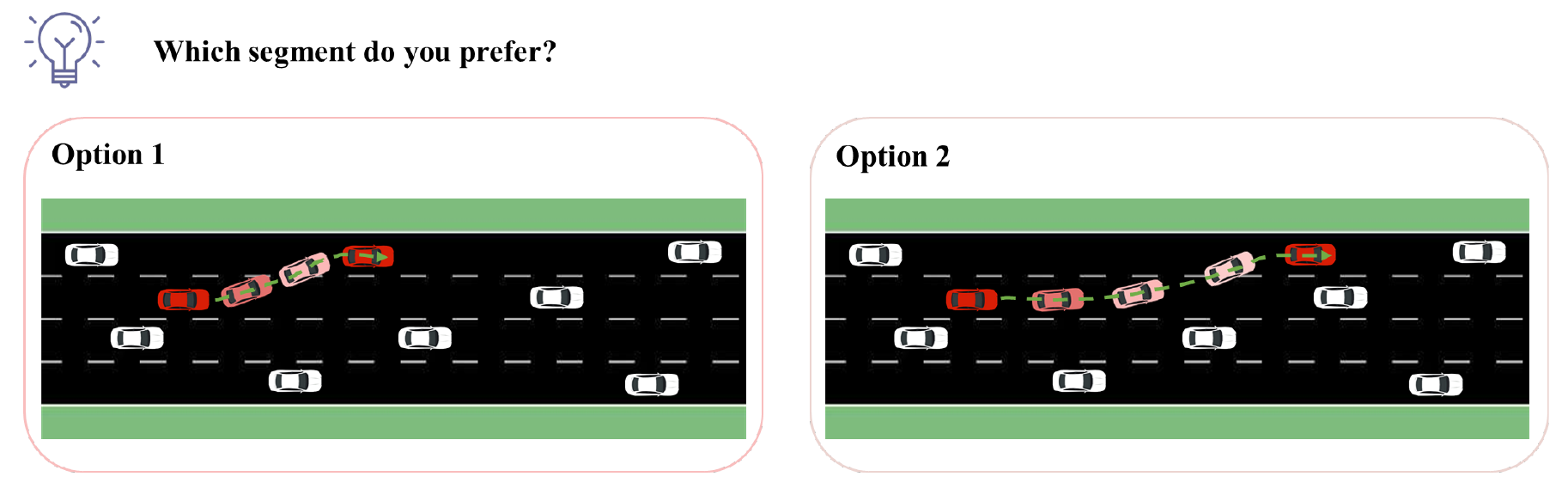}
  \caption{Collection of human feedback using visualized trajectory segments.} 
  \label{fig:fig_option}
\end{figure}

From this generated dataset, we extract pairs of trajectory segments $(\tau^i,\tau^j)$, where $i,j \in \{1,2,\cdots, k\}$ at random. These segments are visualized and presented to human evaluators in a pairwise fashion, as depicted in Figure~\ref{fig:fig_option}. Evaluators are tasked with viewing these clips and indicating a preference for one over the other. In cases where the evaluator perceives both options as equally preferable or cannot establish a preference, they have the option to refrain from labeling that particular pair. This process ensures that the feedback is genuine and reflects a clear preference when present.

The outcomes of these assessments are systematically recorded in a database that captures three key elements: $(x^1,x^2,y)$. Here, $x^1$ and $x^2$ are the time-series data representing the states of the vehicles and lanes over a predefined timestep interval, which correspond to the trajectory segments $\tau^i$ and $\tau^j$, respectively. These states are structured identically to those used during the RL agent's training. The variable $y$ is a binary indicator, taking the form of a distribution over $\{0,1\}$, which denotes the evaluator's preferred segment. When a preference is declared, all weight is assigned to the chosen segment. If the evaluator marks the segments as incomparable, that particular comparison is excluded from the training dataset for the reward model. This methodology ensures that the reward model will be trained on clear human preferences, enhancing the likelihood of developing a policy that truly reflects human-like decision-making in lane-changing scenarios.

\subsubsection{Reward Model Training from Human Assessments}

Upon gathering human preference data, we proceed to fit the reward model $\hat{r}_{\phi}$ to the evaluators' judgments using the collected feedback. We compile a dataset $D=\{(x_i,y_i)_{i=1,...,n}\}$, which consists of examples paired with their corresponding human preferences. The training of parameters $\phi$ is directed towards minimizing the following objective:

\begin{equation} \label{eq:reward_predictor}
    L(D,\phi) = \sum_{i=1}^{n} l(\hat{r}_{\phi}(x_i),y_i)+\lambda_r(\phi),
\end{equation}
where $l$ denotes the cross entropy loss function, and $\lambda_r$ represents a regularization term to prevent overfitting.

\begin{figure}[!ht]
  \centering
  \includegraphics[width=1.0\textwidth]{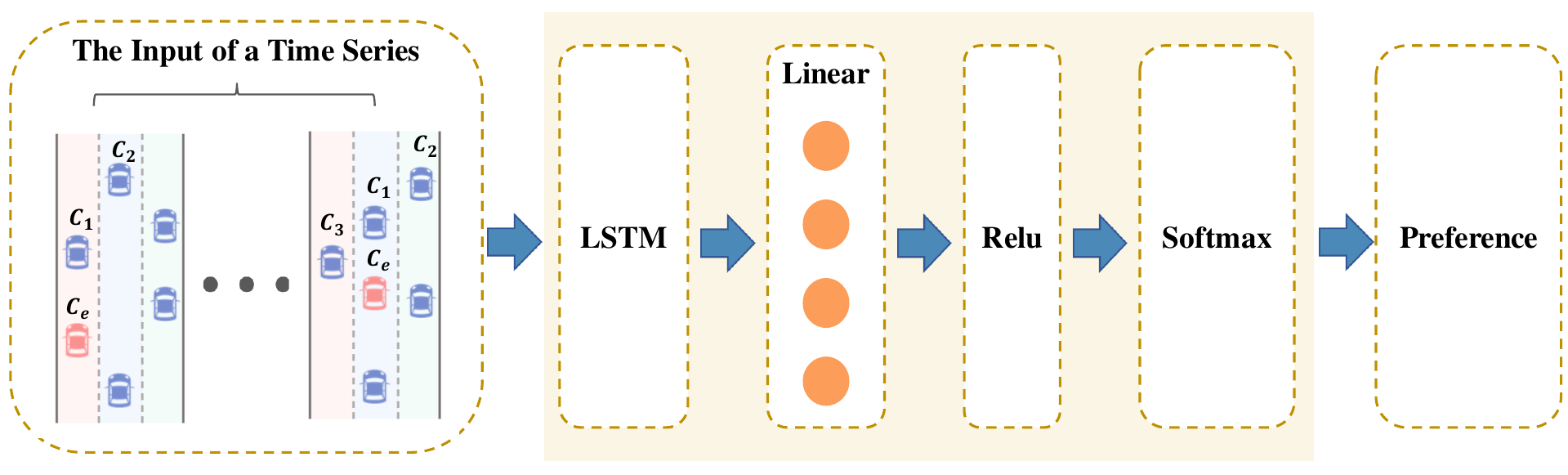}
  \caption{The structure of the reward model.} 
  \label{fig:reward_model}
\end{figure}

The reward model is designed to process temporal sequences that encapsulate essential information about vehicle dynamics and lane configurations. Its core function is to discern human-like behavioral preferences, which serve as the surrogate rewards within the RL framework. Considering that the inference results bifurcate into two categories, \textit{preferred} or \textit{not preferred}, we implement a binary classification model using LSTM networks to approximate the reward function. The LSTM architecture is particularly chosen for its efficacy in capturing temporal dependencies, enabling it to differentiate between datasets characterized by conservative versus aggressive driving behaviors. The overarching objective is to craft a reward model that mirrors human driving preferences, thereby directing the RL agent towards decisions that replicate human lane-changing actions. The structure of the reward model is illustrated in Figure~\ref{fig:reward_model}.

\subsubsection{Policy Refinement for Human-Like Decision Making}

Following the collection of human feedback and the training of the reward model, we proceed to fine-tune the policy to reflect human-like decision-making characteristics. This fine-tuning employs PPO \cite{schulman2017proximal}, which consists of two neural network components: the actor and the critic. The actor network outputs action probabilities based on the current state of the environment, while the critic network estimates the expected returns, essentially assessing the value of the current state. The actor prescribes actions, and the critic evaluates these actions by estimating the value of the resulting states.

PPO uses a policy gradient method to update policies, leveraging reward signals to adjust the likelihood of selecting certain actions, thus promoting beneficial behaviors. The policy gradient update objective is expressed as:

\begin{align} \label{Eq.ppo1}
    L^{PG}(\theta) = \hat{E}_t \left[\log \pi_\theta(a_t | s_t) \hat{A}_t \right],
\end{align}
where $\pi_{\theta}$ denotes the parameterized stochastic policy, and $\hat{A}_t$ is the advantage function estimate at time $t$. The advantage function $\hat{A}_t$ is calculated using the difference between the Q-value (the expected return of the current state-action pair) and the V-value (the expected return of the current state), as follows:

\begin{equation}
    \hat{A}_t = Q(s_t, a_t) - V(s_t).
\end{equation}

The Q-value is typically estimated using the sum of the reward received after taking action $a_t$ in state $s_t$ and the discounted value of subsequent states, while the V-value is estimated by the critic network. These strategies undergo refinement through iterative sampling and optimization of trajectory segments.

Policy gradient methods are particularly sensitive to the choice of step size. A small step size may lead to slow convergence, while a large step size can cause unstable learning trajectories. PPO addresses this challenge by introducing a clipping mechanism in its objective function that constrains updates to the policy, ensuring controlled adjustments between consecutive iterations. The clipped objective is defined as:

\begin{align} \label{Eq.ppo3}
    L^{CLIP}(\theta) = \hat{E}_t \left[\min(r_t(\theta) \hat{A}_t, \text{clip}(r_t(\theta), 1 - \epsilon, 1 + \epsilon) \hat{A}_t) \right],
\end{align}
where $r_t(\theta)$ is the ratio of the probabilities under the new policy to those under the old policy for action $a_t$ in state $s_t$, and $\epsilon$ is a hyperparameter that sets the bounds for clipping, thereby establishing a 'trust region' for policy updates. This clipped objective effectively reduces the risks associated with large policy updates, promoting stability and efficiency in the learning process.

During the fine-tuning phase, human preferences, expressed as rewards, guide the agent's actions. The state and action spaces remain consistent with the pre-training phase, but the reward function is now a reward predictor trained by an LSTM, skilled at capturing diverse preferences. To prevent significant deviations from the pre-trained model, we specifically modify the feature extraction network by constraining the range of parameter updates during training. As illustrated in Figure~\ref{fig:Feature}, we fix the parameters of the linear and LSTM networks, optimizing only the parameters of the last two linear layers. This strategy maintains the generalized performance of the original model while enabling the linear layers' weight parameters to be updated through the training data. As a result, the model better adapts to new environments, comprehends human preferences, and exhibits a driving style akin to that of human drivers.
\section{Experiments} \label{section:experiments}
To accommodate the preferences of conservative and aggressive human drivers, we develop two lane-changing decision models using the RLHF algorithm on the Simulation of Urban Mobility (SUMO) platform. We conduct a series of simulations to compare pre-trained models with both conservative and aggressive models fine-tuned by RLHF, focusing on the ego vehicle's behavior during lane changes. By employing this method, we carry out experiments in scenarios of obstacle avoidance and mixed autonomy, demonstrating RLHF's effectiveness in fine-tuning driving styles.

\subsection{Experimental Details}
All experiments are accomplished on the SUMO platform and we use PyTorch and NVIDIA GeForce RTX 4070 Ti GPUs.
Specifically, we pre-train the lane-changing decision model in both experiments. Based on the current traffic situation, the agent can select from five actions to execute: changing the lane to the left, changing the lane to the right, maintaining the current lane, accelerating at 2 $m/s^2$ acceleration and decelerating at 2 $m/s^2$ deceleration. 
We use Intelligent Driving Models (IDM) to avoid longitudinal collisions between vehicles. 
The goal of pre-training is to allow the vehicles to drive and change lanes basically normally, so we subsequently pay less attention to the acceleration and deceleration control of the vehicles by this model. Besides, we set the lane-changing penalty $r_{lc}$ to -1, and the collision penalty $r_s$ to -10. The efficiency reward $r_e$ is dependent on the road's maximum and minimum speed limits in different scenarios.
The PPO related parameters during pre-training are shown in Table \ref{table:PPO}. 
\begin{table}[!ht]
  \begin{center}
     \caption{Parameters of PPO.}\label{table:PPO}
    \begin{tabular}{cc}
    \toprule
    Parameter name       & Value \\
    \hline
    Batch size  &64\\ 
    Number of steps     &64                                  \\
    Number of epochs    &64                                  \\
    Learning rate       &1e-4          \\
    Clipping range      &	0.2                               \\
    Discount factor     &0.99                                             \\      
    \bottomrule
    \end{tabular}
  \end{center}
\end{table}

We controll the ego vehicle with the pre-trained model and then collect 6,000 sets of visualized trajectory segments of the vehicle's lane-changing maneuvers, extracting the corresponding vehicle and lane information as training data for the reward model. We present the trajectory clips to humans who have a preference for either conservative or aggressive driving styles for labeling. If a person prefers conservative lane-changing behavior, trajectories with an aggressive style are labeled as 0, while those with a conservative style are labeled as 1, and the reverse is true for those who prefer aggressive maneuvers. In this way, we gather human feedback on different preferences. We observe that annotators who favor a conservative style of lane changing prefer to follow vehicles from a longer distance or change lanes when there is ample space ahead. In contrast, those who lean towards an aggressive lane-changing style opt to do so when the gap to the leading vehicle is smaller. Humans also tend to avoid marking scenarios with abnormal lane changes or potential collisions. Particularly, they abstain from labeling segments where the trajectories cannot be comparatively assessed. 

Based on the collected trajectory data and human feedback, we utilize LSTM to fit a reward model. The parameters associated with the reward model are shown in Table \ref{table:reward_predicter}.

\begin{table}[!ht]
  \begin{center}
     \caption{Training parameters of the reward model.}\label{table:reward_predicter}
    \begin{tabular}{ll}
    \toprule
    Parameter Name       & Value \\
    \hline
    Learning rate       & $1 \times 10^{-4}$ \\
    Batch size          & 32 \\
    Weight decay        & $1 \times 10^{-5}$ \\
    Hidden layer size   & 8 \\
    Output size         & 2 \\
    Optimizer           & Adam \\
    \bottomrule
    \end{tabular}
  \end{center}
\end{table}
 
To evaluate the fine-tuning effects of RLHF, we adopt the distance between the RL agent and the preceding vehicle before a lane change as a key metric.
To ascertain the approach for utilizing this metric in evaluation, we conduct a statistical analysis of the data labeled by humans with preferences for different driving styles. We find that aggressive annotators prefer to change lanes when the distance to the vehicle in front is less than 40 meters in more than 85\% of the cases, while conservative annotators show a preference for lane changes when the distance is greater than 40 meters in more than 90\% of scenarios. Consequently, a distance of 40 meters can serve as a threshold to distinguish between conservative and aggressive driving behaviors.
During the specific experimental process, we calculate the proportion of lane changes that occur within a 0-40 meter range from the leading vehicle and those beyond 40 meters, relative to the total number of lane changes, after conducting 1000 simulations for each model.
Additionally, we compare the lateral displacement over time across different models.

\subsection{Simulation in the Obstacle Avoidance}
\subsubsection{Simulation Setting}
\begin{figure}[!ht]
  \centering
  \includegraphics[width=0.8\textwidth]{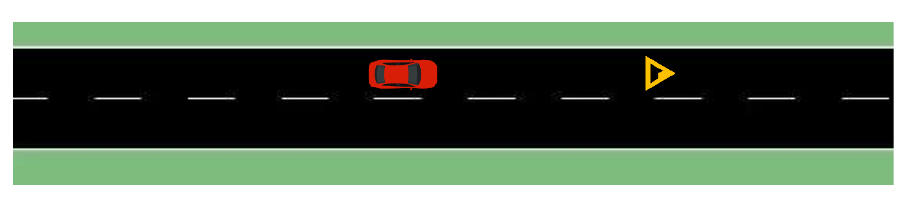}
  \caption{The scenario of the obstacle avoidance.
} \label{fig:fig4}
\end{figure}

We conduct training in the simulation environment as shown in Figure \ref{fig:fig4}, where the ego vehicle could interact with the static object. We select a 600-meter-long urban road with two lanes in each direction. The ego-vehicle enters the road with an initial speed of 40km/h and a maximum speed limit of 70km/h.
On the lane that is 180 meters away from the entrance, we place a stationary vehicle. In each episode, the ego vehicle enters from the lane where the barrier is parked. The vehicle starts the lane change decision after 5s of straight-ahead driving (to facilitate later access to reward models trained by machine learning methods). If a collision occurs between the autonomous vehicle and a stationary vehicle, or if the simulation time exceeds 25s, the episode ends and a new episode begins. The duration of both the visualised trajectory segments and the collected time series is 5 seconds. 

\subsubsection{Evaluation of RLHF's Effectiveness in Fine-tuning Driving Styles}

\begin{figure}[!ht]
  \centering
  \includegraphics[width=0.8\textwidth]{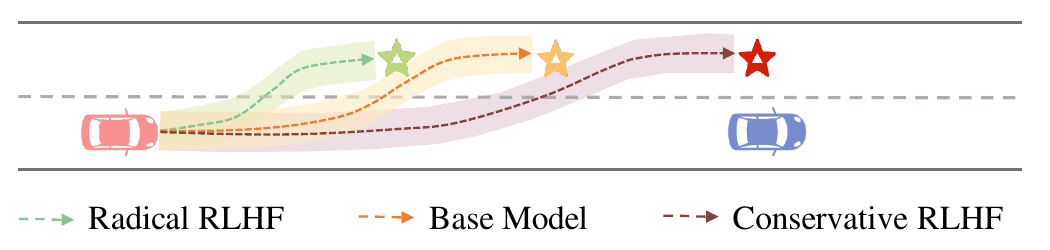}
  \caption{Trajectories of the ego vehicle lane changing process in each model.
} \label{fig:static_result}
\end{figure}

After training the conservative and aggressive RLHF models with different preferred reward functions, we conduct 1000 simulation runs, each lasting 25 seconds, to collect comparative driving data.
As depicted in Figure \ref{fig:static_result}, from the visual interface observed during the intuitive experimental process, we can see that the conservative RLHF tends to change lanes when there is a greater distance from the leading vehicle, while the aggressive RLHF prefers to change lanes when closer to the leading vehicle.

\begin{figure}[!ht]
    \centering
    \subfigure[Statistical results of the distance between the ego vehicle and the leading vehicle in the same lane prior to a lane change]{\includegraphics[width=0.47\linewidth]{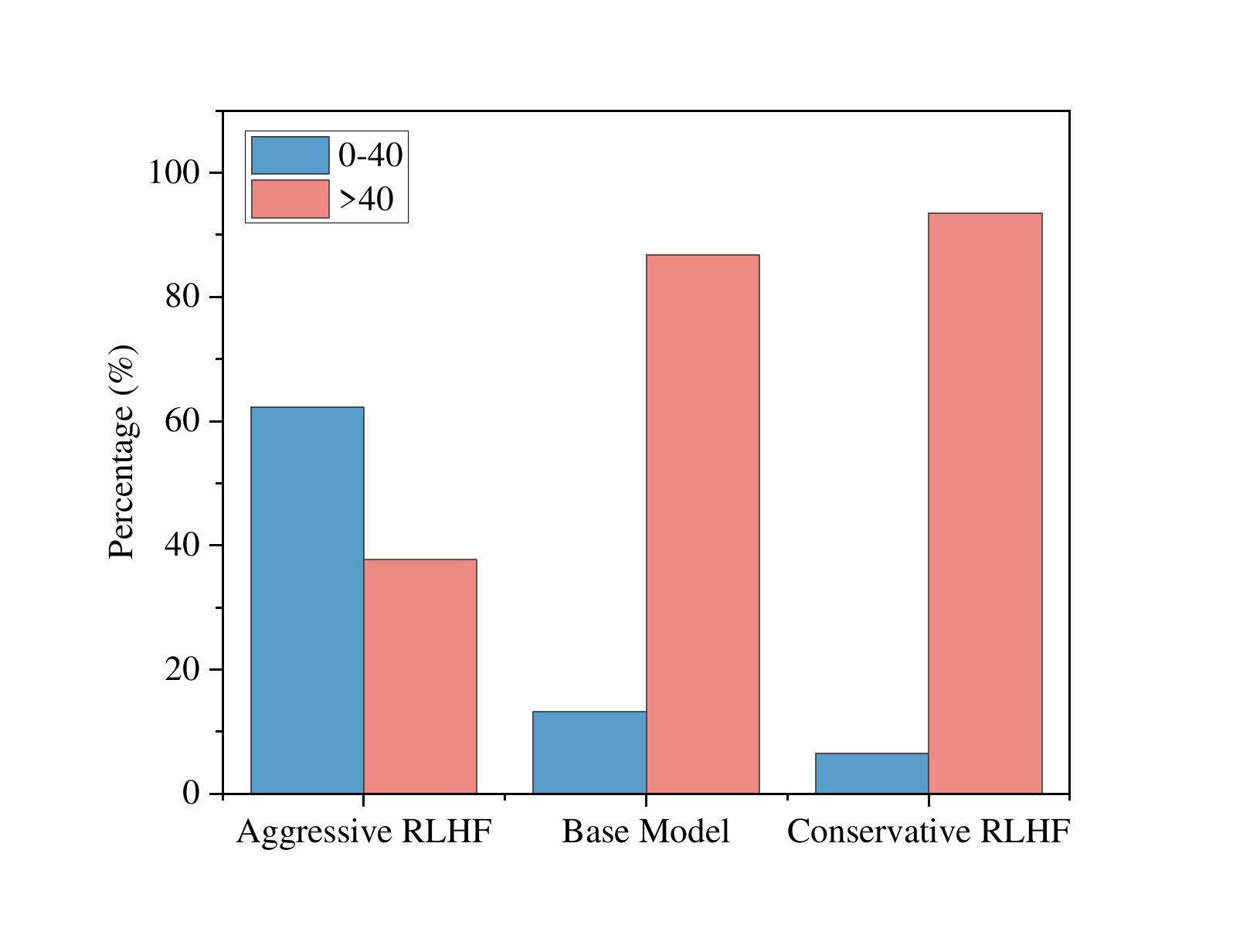}}
    \hspace{5mm}
    \subfigure[The lateral position of the ego vehicle over time during one episode of simulation with the last checkpoint of the three types of models]{\includegraphics[width=0.47\linewidth]{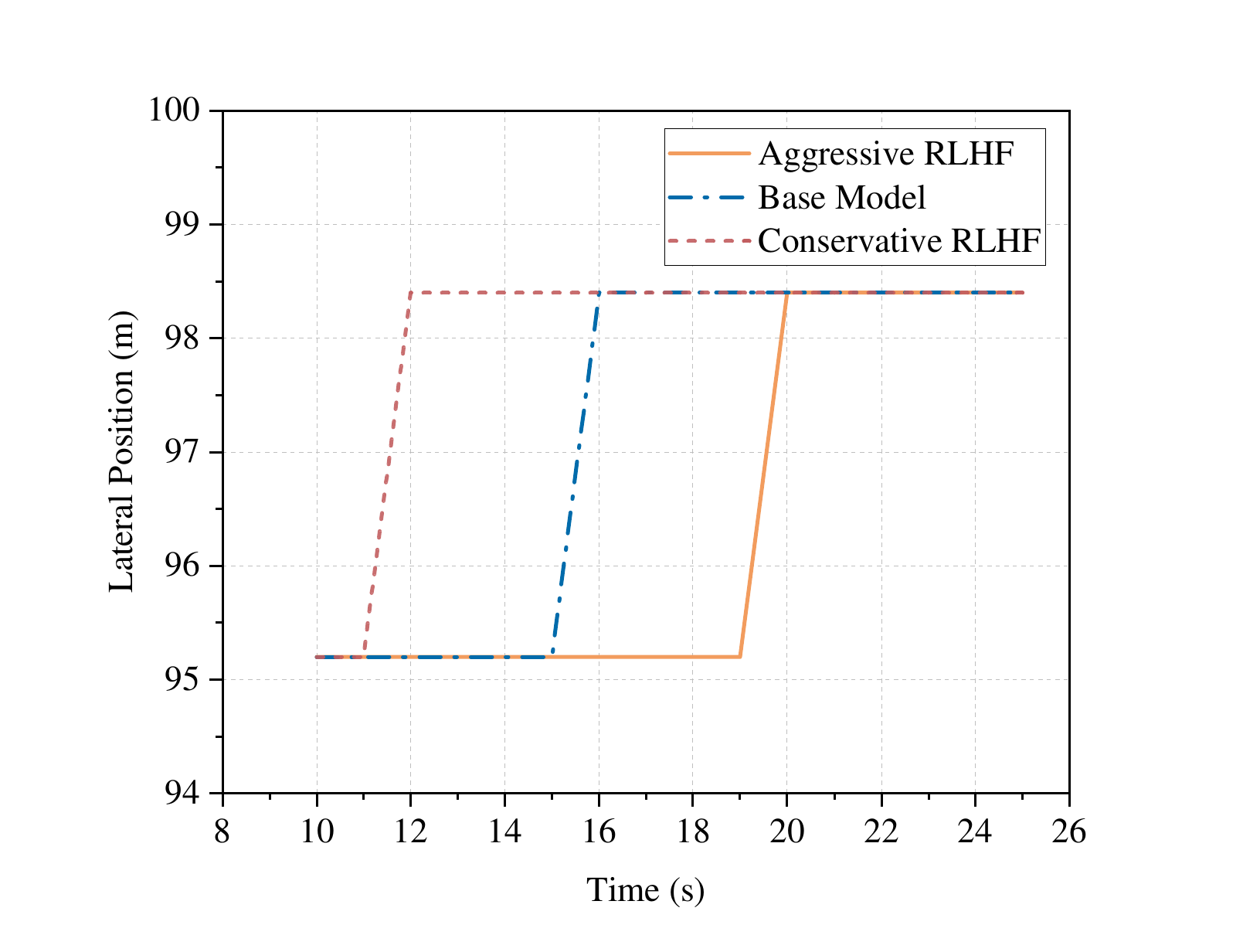}}
    \caption{Results of style fine-tuning for the RLHF model in obstacle avoidance.}
    \label{Figure 9}
\end{figure}

As shown in Figure \ref{Figure 9}(a), the conservative RLHF model shows the smallest proportion of lane changes within a distance of 0-40m, with the pre-trained model in an intermediate position, and the aggressive RLHF model shows the largest proportion. Conversely, when lane changing distances beyond 40m are considered, the conservative RLHF model has the largest proportion, while the aggressive RLHF model has the smallest. 
Compared to the base model, the aggressive RLHF model showed a 49.1\% increase in the number of lane changes within the critical range of 0-40 metres and the conservative RLHF model showed a 6.7\% increase in the number of lane changes beyond 40 metres.
As shown in Figure~\ref{Figure 9}(b), we operate the ego vehicle using the last checkpoint of each model. The result indicates that the aggressive RLHF model tends to initiate lane changes at a later time compared to other models, in contrast to the conservative RLHF which makes lane changes earliest.

\subsection{Simulation in the Mixed Autonomy}

\subsubsection{Simulation Setting}

\begin{figure}[!ht]
  \centering
  \includegraphics[width=0.8\textwidth]{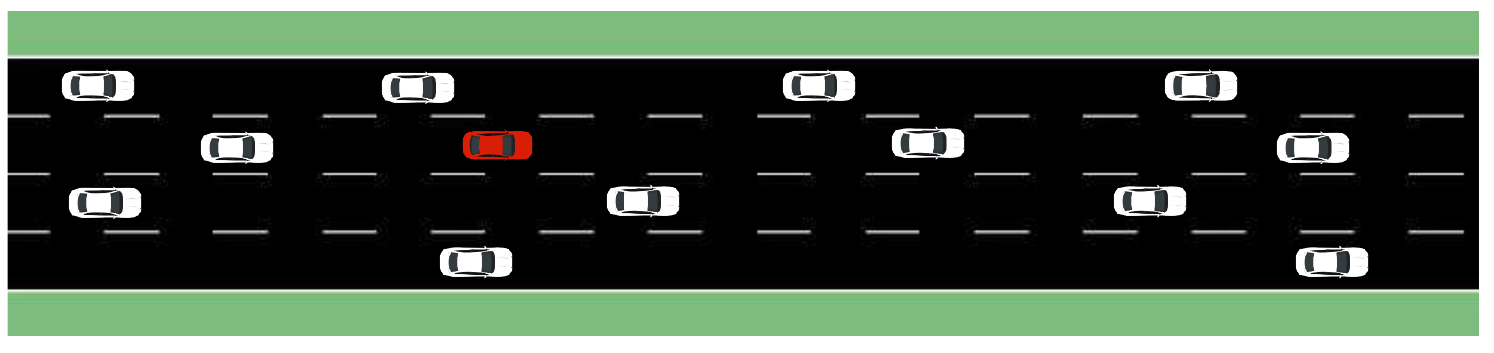}
  \caption{Scenario of the mixed autonomy.
} \label{fig:SUMO}
\end{figure}

We enable AVs to interact with multiple moving vehicles in a simulation environment as shown in Figure \ref{fig:SUMO}. The simulation environment is set up as follows: the road is 1000m long, each lane in the road is 3.75m wide, and the speed limit of each lane is 120km/h. The environment is set up with one autonomous vehicle and several human driven vehicles, with the length of the vehicles set to 5m. To ensure that the training vehicle has sufficient motivation and space to change lanes, the arrival time and corresponding lanes of the training and social vehicles are artificially set. In addition, the maximum speed of the training vehicle is set to 120km/h, the minimum speed is set to 80km/h and the maximum speed of the social vehicle is set to 72km/h. All vehicles except the ego vehicle drive in their current lane. In addition, the duration of both the visualised trajectory segments and the collected time series was 8s.

\subsubsection{Evaluation of RLHF's Effectiveness in Fine-tuning Driving Styles}

\begin{figure}[!ht]
    \centering
    \subfigure[Statistical results of the distance between the ego vehicle and the leading vehicle in the same lane prior to a lane change]{\includegraphics[width=0.47\linewidth]{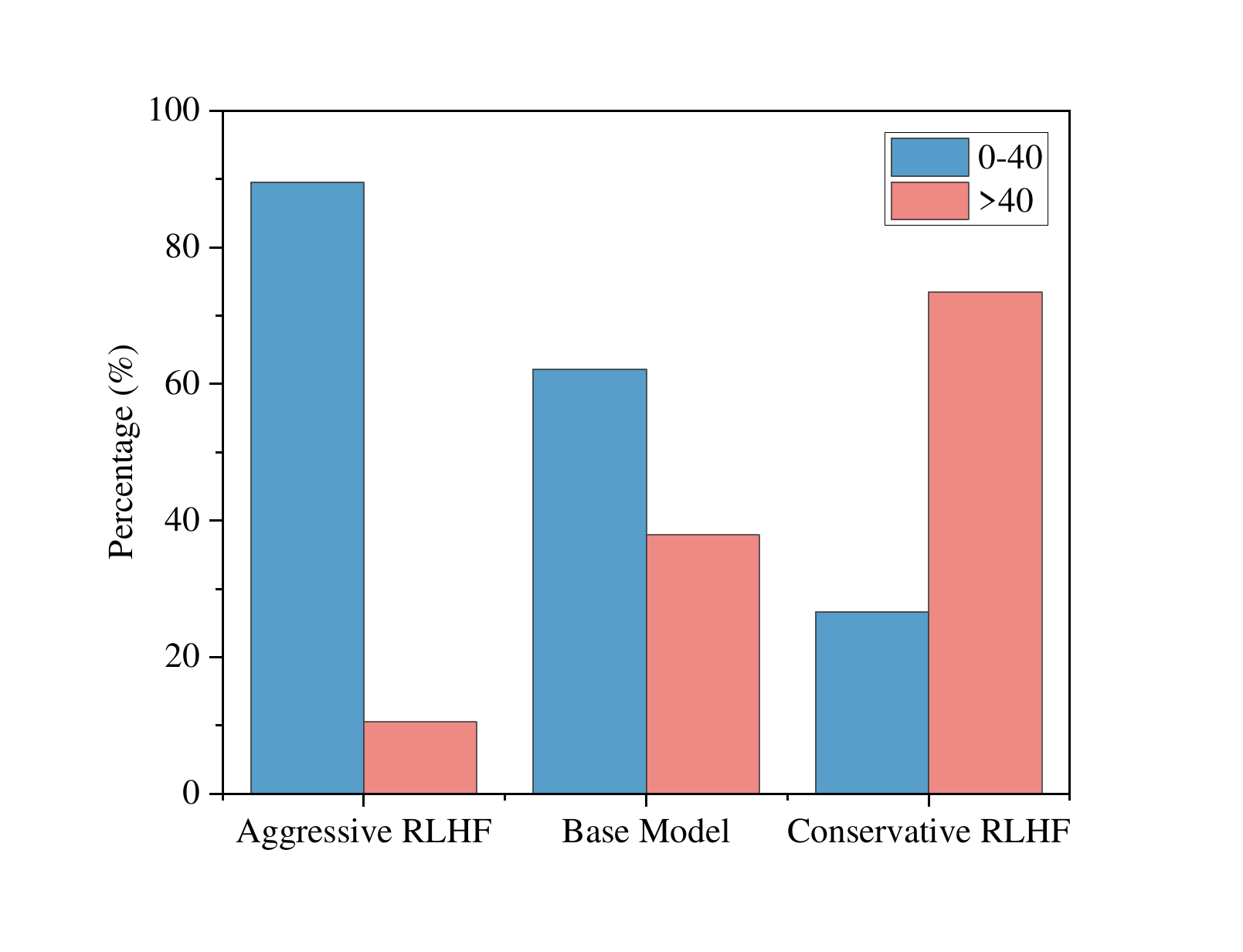}}
    \hspace{5mm}
     \subfigure[The lateral position of the ego vehicle over time during one episode of simulation with the last checkpoint of the three types of models]{\includegraphics[width=0.47\linewidth]{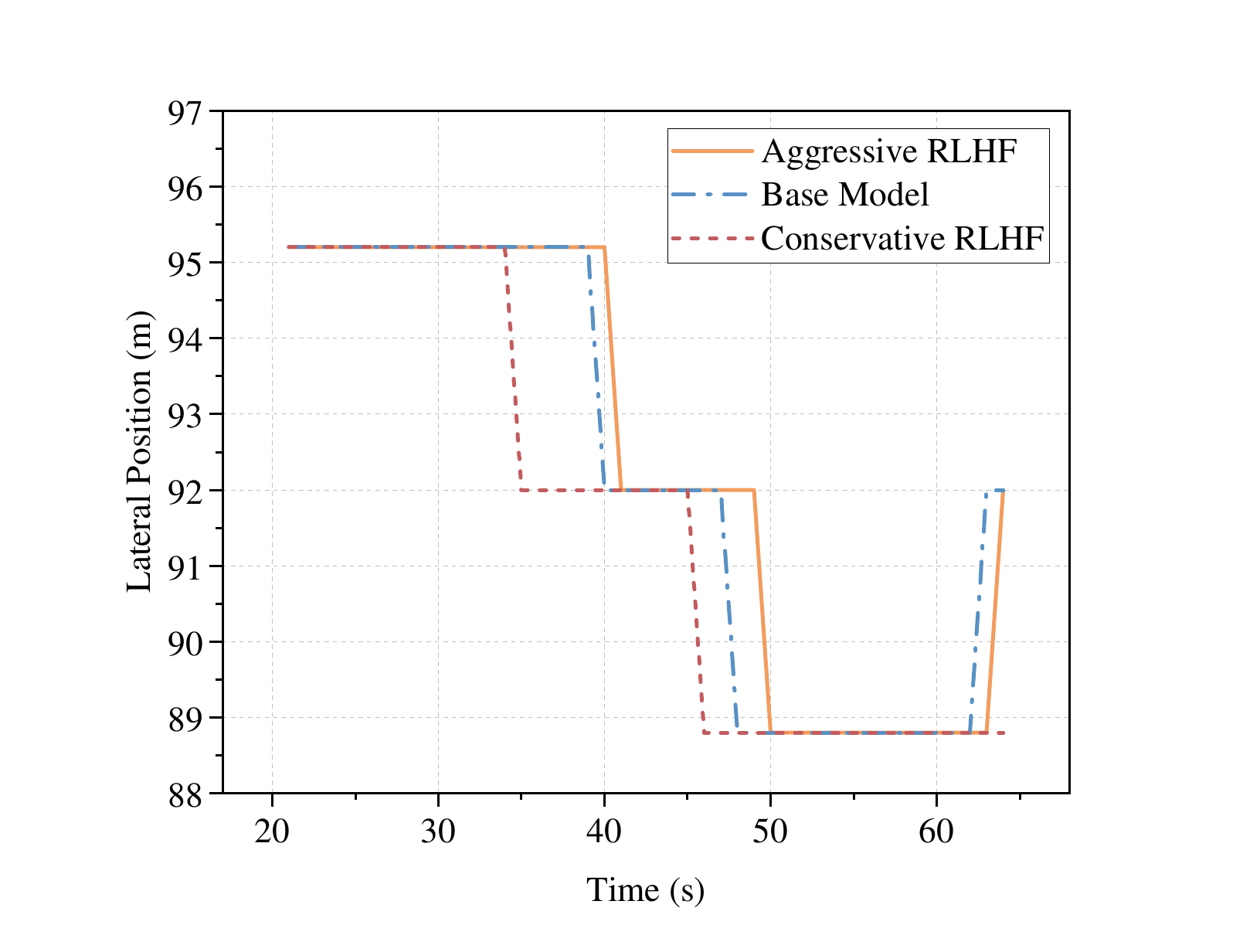}}
    \caption{Results of style fine-tuning for the RLHF model in mixed autonomy.}
    \label{Figure 13}
\end{figure}

We use three different models to control the decision making of the autonomous vehicle in the mixed autonomy. We compare metrics of the ego vehicle controlled by the base model, aggressive RLHF model, and conservative RLHF model, such as the distance to the leading vehicle prior to a lane change. As shown in Figure \ref{Figure 13}(a), compared with the base model, the aggressive RLHF model increases the number of lane changes from the vehicle in front by 27.4\% in the 0-40 interval and the conservative RLHF model increases the number by 35.5\% when the distance is greater than 40 meters. 
Furthermore, as shown in Figure \ref{Figure 13}(b), we find that the lane-changing time of the aggressive RLHF is later than that of the other models, and the aggressive RLHF model prefers lane-changing more than conservative RLHF model. 

\begin{figure}[!ht]
  \centering
  \includegraphics[width=0.8\textwidth]{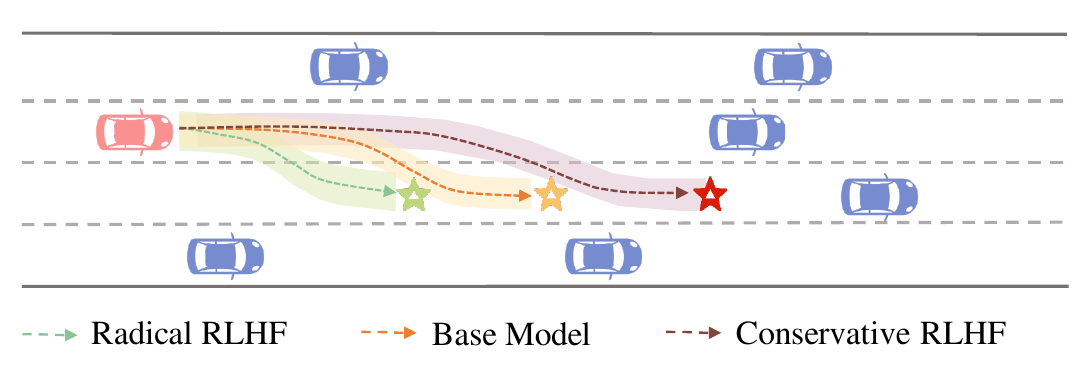}
  \caption{Trajectories of the ego vehicle lane changing process in each model.
} \label{fig:dynamic_result}
\end{figure}

It can also be seen from the visualization in Figure \ref{fig:dynamic_result} that the aggressive RLHF model changes lanes in most cases when it is closer to the vehicle in front, while the conservative RLHF model changes lanes when it is farther away from the vehicle in front. The evaluation results we obtain are similar to those obtained in the experiment of the previous section, again validating the effectiveness of the RLHF model in improving the human-like decision making of autonomous driving.
\section{Conclusions} \label{section:conclusion}
This article proposes a model for lane-changing decisions of vehicles in road networks. We use RLHF to understand and model real human preferences, which are used to fit reward models to guide the training of RL. Our goal is to realize human-like lane changing decision making for AVs. We use the SUMO platform to experiment with two different scenarios, the obstacle avoidance and the mixed autonomy. Based on the pre-trained lane-changing model, we collect conservative and aggressive human labeling data, train the reward models using the LSTM algorithm, and then fine-tune them by PPO algorithm to get the conservative and aggressive RLHF models respectively. The effectiveness of the RLHF method is verified by the different preferences of vehicle lane changing decision-making styles in the experimental results.

This work can be extended in several directions. First, consider further testing the effectiveness of the algorithm in real vehicle trials. Second, extend RLHF to other complex scenarios, such as the decision-making of AVs at intersections. Third, extend the model to the field of multi-agent interactions.

\newpage

\bibliographystyle{trb}
\bibliography{trb_template}
\end{document}